\documentclass[aip,reprint,apl,amssymb,citesort]{revtex4-1}
\usepackage{amsmath}
\usepackage{graphicx}
\usepackage{color}
\begin{document}

\title{Low thermal conductivity and triaxial phononic anisotropy of SnSe}

\author{Jes\'us Carrete}
\author{Natalio Mingo}
\email{natalio.mingo@cea.fr}
\affiliation{CEA-Grenoble, 17 Rue des Martyrs, Grenoble 38054, France}

\author{Stefano Curtarolo}
\email{stefano@duke.edu}
\affiliation{Center for Materials Genomics, Materials Science, Electrical Engineering, Physics and Chemistry, Duke University, Durham, North Carolina 27708, USA}

\begin{abstract}
  In this theoretical study, we investigate the origins of the very low thermal conductivity of tin selenide (SnSe) using \textit{ab-initio} calculations. We obtained high-temperature lattice thermal conductivity values that are close to those of amorphous compounds. We also found a strong anisotropy between the three crystallographic axes: one of the in-plane directions conducts heat much more easily than the other. Our results are compatible with most of the experimental literature on SnSe, and differ markedly from the more isotropic values reported by a recent study.
\end{abstract}

\maketitle

The ongoing quest for improved thermoelectric materials has spanned the past six decades.\cite{goldsmid,nmatHT} Good thermoelectrics are characterized by their high figures of merit $ZT=PT/\kappa$. This requires a low thermal conductivity $\kappa$, and a high power factor $P=\sigma S^2$ (where $\sigma$ is the electrical conductivity, and $S$ the thermopower) across their intended range of working temperatures. Nanotechnology has been able to greatly improve on earlier single-crystal thermoelectric materials \cite{Minnich2009} via hindering phonon flow while keeping favorable electric properties, even when starting from a poor performer such as bulk silicon.\cite{bux_nanostructured_2009}

Thus, it was somewhat surprising when single-crystal SnSe recently emerged as the best thermoelectric material ever measured.
  Experimental measurements on monocrystalline samples\cite{Kanatzidis_Nature_SnSe_2014} point to a record figure of merit of $ZT = 2.6$ at $T = 923\,\mathrm{K}$.
   This is due mainly to its very low lattice thermal conductivity $\kappa_{\ell}$.
  Intense interest in SnSe was spurred by these results, and to date two independent sets of measurements on polycrystals have been published.\cite{Lenoir_APL_SnSe_2014,Snyder_JMCA_SnSe_2014}
  The two series are compatible with each other, yet they both display higher values of $\kappa$ than those reported in Ref. \onlinecite{Kanatzidis_Nature_SnSe_2014} for the single crystal.
  This contrasts with the expectation that grain boundaries should  decrease the thermal conductivity below that of the single crystal.
  Remarkably, preexisting studies of single crystals \cite{Wasscher_maxZT_SnS_SnSe_SSE_1963} reported even higher values of $\kappa$, with a directional maximum of around  $1.8\,\mathrm{W\,m^{-1}\,K^{-1}}$ at room temperature.

To address this controversy, we studied phonon transport in SnSe from first principles based on the Boltzmann transport equation (BTE) formalism.
  Such modeling studies offer substantially more detailed information than is readily available from the experiments, and have demonstrated both their wide range of applicability and their predictive power.\cite{Broido_APL_kappa_2007,ward_intrinsic_2010,Dong_MgO_kappa_PNAS_2010,Li_thermal_2013, Lindsay_phonon_kappa_PRB_2014}  
For the present work, we focused on the Pnma-symmetric phase of SnSe.
  This is the stable phase from room temperature up to $\sim 750-800\,\mathrm{K}$.\cite{Kanatzidis_Nature_SnSe_2014}
  Hence, it is the pertinent phase for most of the aforementioned measurements.\cite{Kanatzidis_Nature_SnSe_2014,Lenoir_APL_SnSe_2014,Snyder_JMCA_SnSe_2014}

\begin{figure}
  \begin{center}
    \includegraphics[width=0.9\columnwidth]{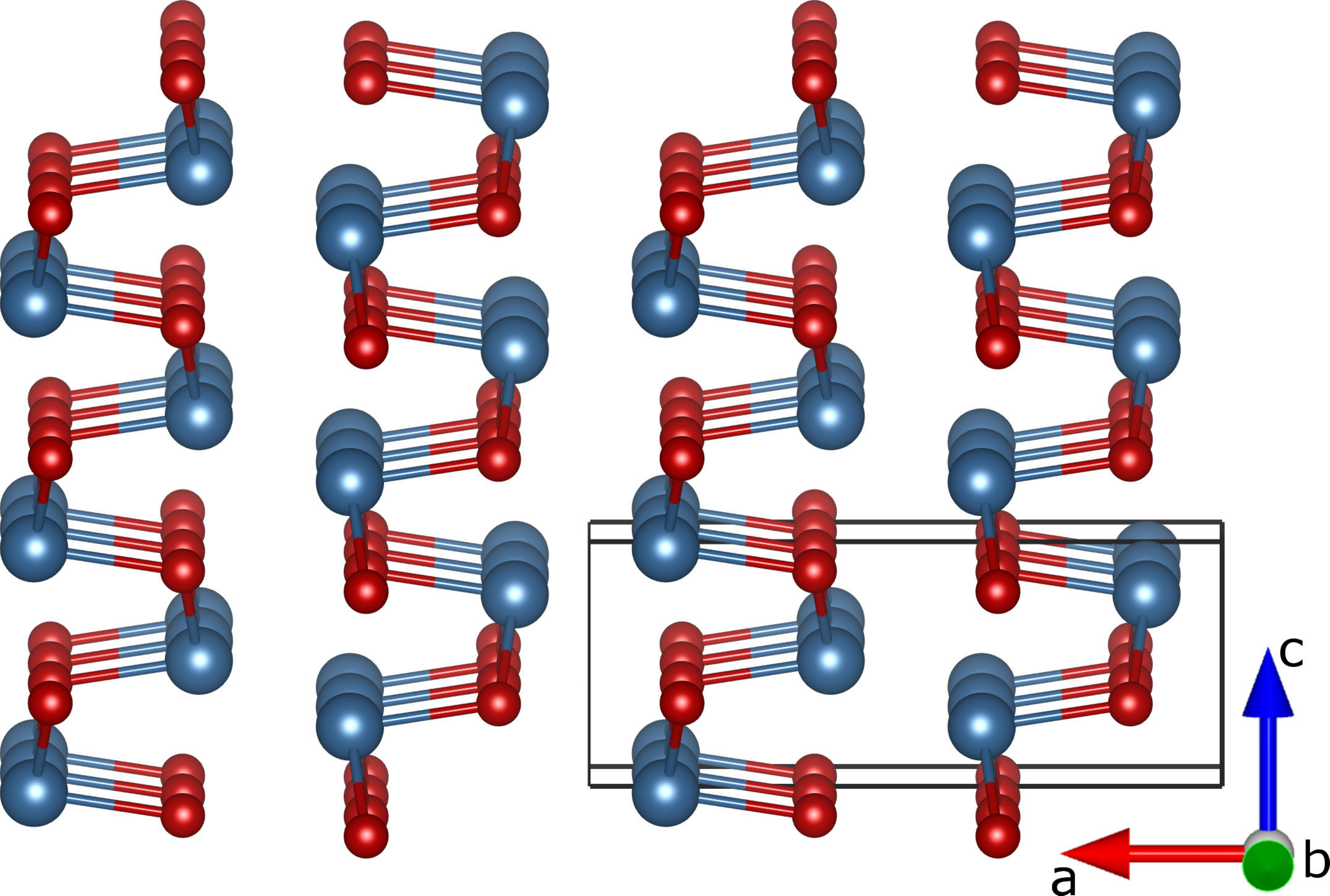}
  \end{center}
  \vspace{-5mm}
  \caption{View of a $2\times 3\times 3$ supercell of Pnma SnSe. A grey prism outlines a single unit cell, containing $8$ atoms.}
  \label{fig:view}
\end{figure}

  We started with an atomistic description of Pnma SnSe extracted from the {\small AFLOWLIB}.org consortium repository [auid=aflow:d188eb9b8df60e58].\cite{aflowlibPAPER,aflowAPI} 
  We relaxed the structure without any geometrical constraint using the density functional theory (DFT) software package VASP.\cite{vasp} All technical details are presented in the supplementary material.
  We obtained unit cell lengths of  $a=11.72\,\mathrm{\AA}$, $b=4.20\,\mathrm{\AA}$ and $c=4.55\,\mathrm{\AA}$, which are in reasonable agreement with the literature.\cite{VonSchnering_neutron_SnS_SnSe_JPCS_1986,Kanatzidis_Nature_SnSe_2014,Lenoir_APL_SnSe_2014}
  The relaxed structure is shown in Fig. \ref{fig:view}. From a geometric point of view, this phase is quasi-laminar: the ``in-plane'' $b$ and $c$ axes are almost equivalent, while the ``cross-plane'' $a$ axis is  triple their length. 

We next computed sets of second- and third-order interatomic force constants using a $3\times5\times5$ supercell containing $600$ atoms. We additionally obtained the dielectric parameters of the system to account for long-range Coulombic interactions: a set of Born effective charges, and the dielectric tensor in the infinite-frequency limit. This information is sufficient for obtaining the phonon spectrum and three-phonon scattering rates required to solve the  BTE. We used the Phonopy package \cite{phonopy} for the second-order calculations. To obtain the third-order force constants and to solve the BTE, we employed our own software, ShengBTE,\cite{ShengBTE_2014} based on an iterative solution method\cite{Omini_iterative_boltzmann_PhysicaA_1995} and a locally adaptive smearing scheme\cite{Wu_PRB_2012} to enforce energy conservation. All the required forces were obtained from DFT calculations with VASP. Our workflow is documented in full detail elsewhere, and the source code we used in this implementation is publicly available for download under an open source license.\cite{ShengBTE_2014}

\begin{figure}
  \begin{center}
    \includegraphics[width=\columnwidth]{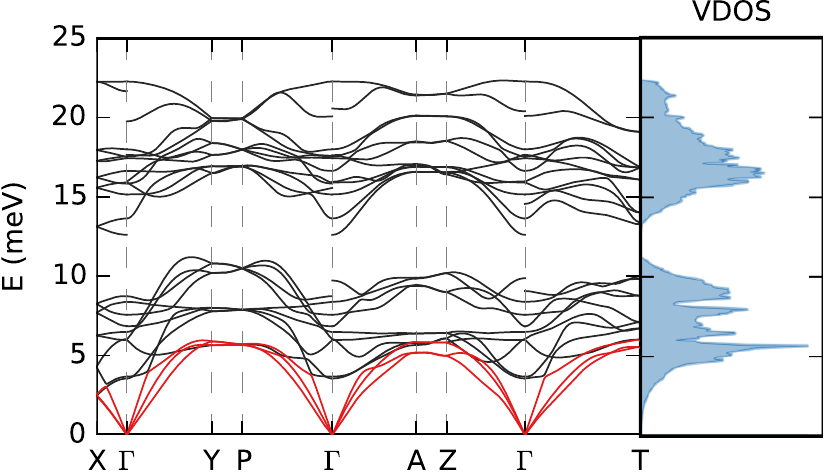}
  \end{center}
  \vspace{-5mm}
  \caption{Phonon dispersion curves and vibrational density of states (VDOS) for the Pnma phase of SnSe. Acoustic modes are highlighted in red.}
  \label{fig:spectrum}
\end{figure}

The computed phonon spectrum is shown in Fig. \ref{fig:spectrum}.
  The vibrational density of states has a bipartite structure, with a gap around $11.8\,\mathrm{meV}$ that delimits two groups, each with twelve branches.
  The Pnma unit cell of SnSe may be conceptualized as a distorted variation of a simpler diatomic rocksalt structure.\cite{Kanatzidis_Nature_SnSe_2014}
  This is consistent with a bipartite spectrum whose lower (upper) half corresponds roughly to the folded acoustic (optical) modes of the rocksalt unit cell.
  Another conclusion that may be drawn from this line of reasoning is that speeds of sound  should be roughly isotropic.
  This is confirmed by direct calculation, and is also apparent in Fig. \ref{fig:spectrum}.
  Nevertheless, the much shorter length of the reciprocal-space unit cell along $a^{*}$ allows us to anticipate a much lower value of $\kappa_{\ell}$ along $a$.
  This lowered value arises from purely geometric considerations.

\begin{figure}
  \begin{center}
    \includegraphics[width=\columnwidth]{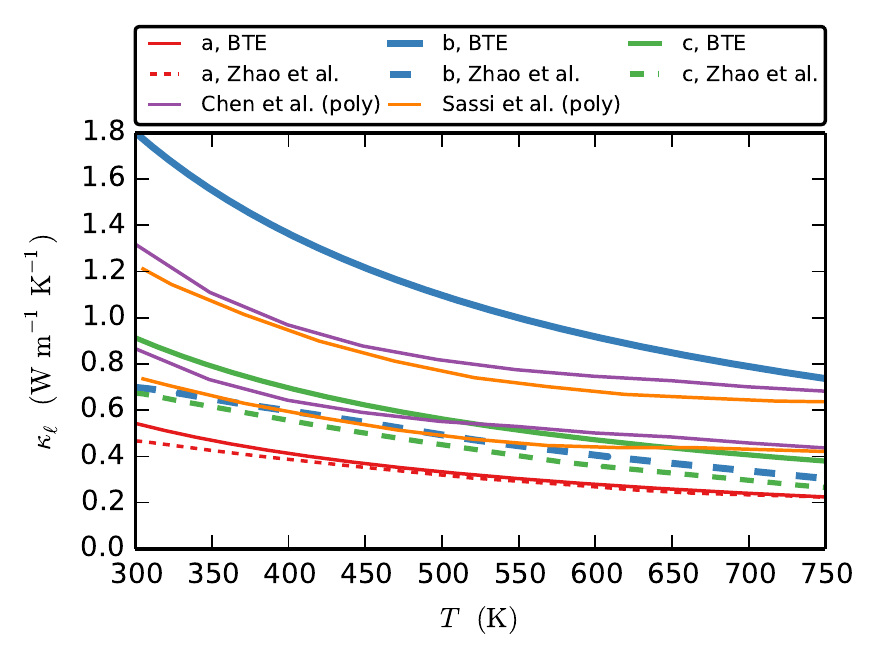}
  \end{center}
  \vspace{-5mm}
  \caption{Lattice thermal conductivity of the Pnma phase of SnSe: comparison between experimental results  for monocrystals\cite{Kanatzidis_Nature_SnSe_2014} and polycristals,\cite{Lenoir_APL_SnSe_2014,Snyder_JMCA_SnSe_2014} and theoretical computations in this letter.}
  \label{fig:kappa}
\end{figure}

Our results for single-crystal SnSe, at temperatures from $300\,\mathrm{K}$ to $775\,\mathrm{K}$, are plotted and compared with experimental data \cite{Kanatzidis_Nature_SnSe_2014} in Fig. \ref{fig:kappa}. 
All values of $\kappa_{\ell}$ are remarkably low for a crystalline material.\cite{Goodson_min_kappa_Science_2007} This is partly due to the large number of optical modes and the low frequencies in the spectrum. Low frequencies are the product of a combination of heavy constituent elements and low atomic coordination. The influence of coordination is evidenced by the fact that the typical energies of phonons below the gap are comparable to those of acoustic phonons in rocksalt PbTe\cite{Cochran_crystal_PbTe_1966} despite the latter's much heavier elements. Furthermore, we have also calculated the phonon spectrum of the rocksalt phase of SnSe\footnote{See supplementary material at http://dx.doi.org/} [auid=aflow:c14b4f7b1a88de19], and found that the acoustic modes of rocksalt SnSe have higher frequencies than those of PbTe, as could be predicted considering their masses. As regards the influence of optical modes, even though they make up almost $90\%$ of the total number of modes, they only account for $\sim 45\%$ of $\kappa_{\ell}$ along $a$, and $\sim 65\%$ of its values along $b$ or $c$. Their low average group velocities are responsible for this underrepresentation. Other systems with complex unit cells and many optical modes are commonly associated with low thermal conductivities.\cite{slack}

  We detected a strong anisotropy between the three axes, with $\kappa_{\ell}^{b}>\kappa_{\ell}^{c}>\kappa_{\ell}^{a}$ at all temperatures.
  Our computed value of $\kappa_{\ell}^{a}$ is in rather good agreement with experimental measurements.
  Intriguingly, our results for the remaining two axes disagree markedly with Ref. \onlinecite{Kanatzidis_Nature_SnSe_2014}.
  The recent experimental results point to an almost isotropic thermal conduction along the $bc$ plane, and an in-plane room-temperature $\kappa_{\ell}\sim 0.7\,\mathrm{W\,m^{-1}\,K^{-1}}$.
  Instead, we clearly identify $b$ as a much more conductive axis for phonons, with a factor-of-two difference in $\kappa_{\ell}$ with respect to $c$, and an average in-plane $\kappa_{\ell}\sim 1.4\,\mathrm{W\,m^{-1}\,K^{-1}}$ at $300\,\mathrm{K}$.
  Our data are more in line with historical measurements on this system.\cite{Wasscher_maxZT_SnS_SnSe_SSE_1963}
  The origin of this $bc$ anisotropy can be traced to the phonon dispersions in Fig. \ref{fig:spectrum}: the average phonon frequencies and group velocities are higher along $\Gamma\rightarrow Y$ than along $\Gamma\rightarrow Z$.
  A finer analysis reveals that the source of this in-plane anisotropy are modes in the lower half of the spectrum --- in fact, the contribution to $\kappa_{\ell}$ from optical modes above the gap is very similar along both directions. Along $c$, optical modes below and above the gap are almost on equal footing, whereas along $b$ the lower-frequency group is responsible for more than $70\%$ of the total optical contribution.

To test the hypothesis that the frequencies and group velocities of modes below the gap give rise to the anisotropic in-plane conduction, we have studied the tensor quantity:
\begin{equation}
C^{\alpha\beta}=\frac{24k_B}{\Omega}\left[\frac{x}{\sinh \left(x\right)}\right]^2v^{\alpha}v^{\beta},
\label{eqn:smalltau}
\end{equation}
\noindent for each branch at each point of the Brillouin zone (BZ).
Here, $\Omega$ is the unit cell volume, $k_B$ is the Boltzmann constant, $x$ is defined as $E/\left(2k_B T\right)$, $E$ is a vibrational energy and $\mathbf{v}$ a group velocity. 
  The average of $C^{\alpha\beta}\tau$ (where $\tau^{-1}$ is the scattering rate) among phonon branches and over the BZ is the lattice thermal conductivity tensor.\cite{ShengBTE_2014} 
Thus, $C^{\alpha\beta}$ accounts for the influence of the phonon spectrum on $\kappa_{\ell}$. 
For the phonon branches below the gap in SnSe, $C^{bb}$ is on average two to three times higher than $C^{cc}$. 
As scattering rates are substantially higher for the high-frequency modes, it is modes below the gap that contribute the most significantly to $\kappa_{\ell}$.
  Hence, the anisotropy of $C^{\alpha\beta}$, determined by frequencies and group velocities alone, is sufficient to explain the difference between $\kappa_{\ell}^b$ and $\kappa_{\ell}^c$.

As the phonon spectrum is computed from forces caused by small nuclear displacements, the anisotropy of $C^{\alpha\beta}$ must ultimately be due to a mechanical anisotropy of the material.
  As is illustrated in Fig. \ref{fig:view}, despite the similar dimensions of its unit cell along $b$ and $c$, Pnma SnSe is far from isotropic in the $bc$ plane.
  Along $b$, the compound can be considered as a series of zig-zaging SnSe chains, with a Sn-Se distance of $2.82\,\mathrm{\AA}$.
  Along $c$, a second Sn-Se distance of $3.43\,\mathrm{\AA}$ must also be considered.

Our analysis so far has been confined to the direct influence of the phonon spectrum on the thermal conductivity. The question remains, however, if the strength of phonon-phonon interaction in Pnma SnSe also plays a role in explaining its ultra-low thermal conductivity. To answer it, we have computed a representative averaged value of the phonon scattering rate for each axis by dividing the appropriate diagonal component of the tensor defined by \eqref{eqn:smalltau} over the corresponding principal thermal conductivity. Note that, although scattering rates are scalars, a directional dependence is introduced by the weighting of each mode in the average.  At $T=300\,\mathrm{K}$, the results are $0.17$, $0.23$ and $0.28\,\mathrm{ps^{-1}}$ for the $a$, $b$ and $c$ axes respectively. By way of comparison, the same quantity takes the value $0.031\,\mathrm{ps^{-1}}$ for crystalline Si at the same temperature when calculated using the data in Ref. \onlinecite{ShengBTE_2014}. Interestingly, the representative scattering rate along $a$ is the lowest despite its being the lowest-thermal-conductivity axis. This further supports the idea that the origin of the cross-plane anisotropy is geometric, as pointed out above.

\begin{figure}
  \begin{center}
    \includegraphics[width=\columnwidth]{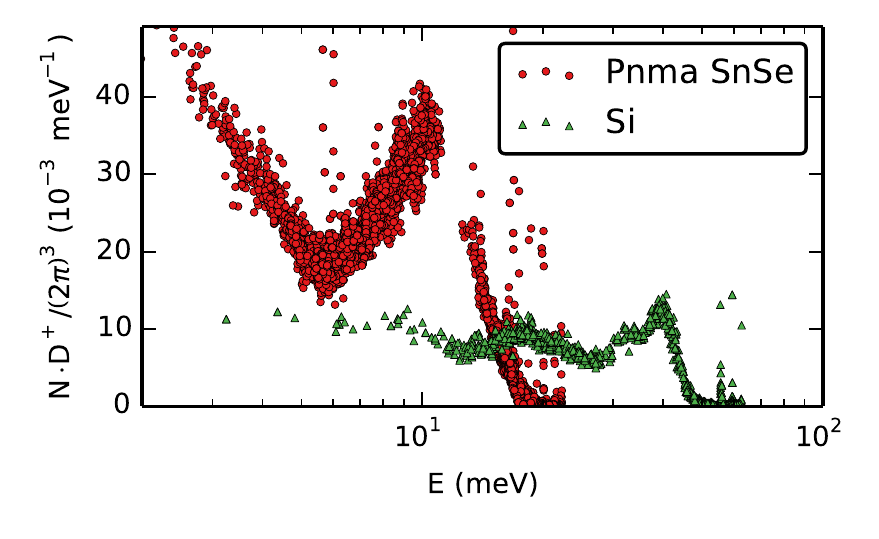}
  \end{center}
  \vspace{-5mm}
  \caption{Contribution of each vibrational mode to the anharmonic phase space volume of silicon and Pnma SnSe. $4\times 12\times 12$ and $16\times 16\times 16$ and $20\times 20\times 20$ meshes were used to sample reciprocal space, respectively. $N$ denotes the total number of points sampled.}
  \label{fig:P3}
\end{figure}

Such high scattering rates can be due to the very anharmonic nature of the crystal or to a large number of allowed three-phonon processes. To quantify both factors we have computed the mode Gr\"uneisen parameters and anharmonic phase space volume $P_3$,\cite{Lindsay_JPCM_2008} respectively. The former yield a high-temperature-limit total Gr\"uneisen parameter $\gamma=0.63$, only $\sim 10\%$ higher than that of Si.\cite{Balandin_qwell_kappa_PRB_1998} In contrast, $P_3=1.9\cdot 10^{-2}\,\mathrm{eV^{-1}}$, very high in comparison with Si ($P_3=0.47\cdot 10^{-2}\,\mathrm{eV^{-1}}$) or even CdTe ($P_3=1.42\cdot 10^{-2}\,\mathrm{eV^{-1}}$),\cite{Lindsay_JPCM_2008} another known low-conductivity compound. Hence, the number of allowed three-phonon processes is the cause of the enhanced scattering. Figure \ref{fig:P3} shows the comparison between the phase spaces of Pnma SnSe and Si in detail.

\begin{figure}
  \begin{center}
    \includegraphics[width=\columnwidth]{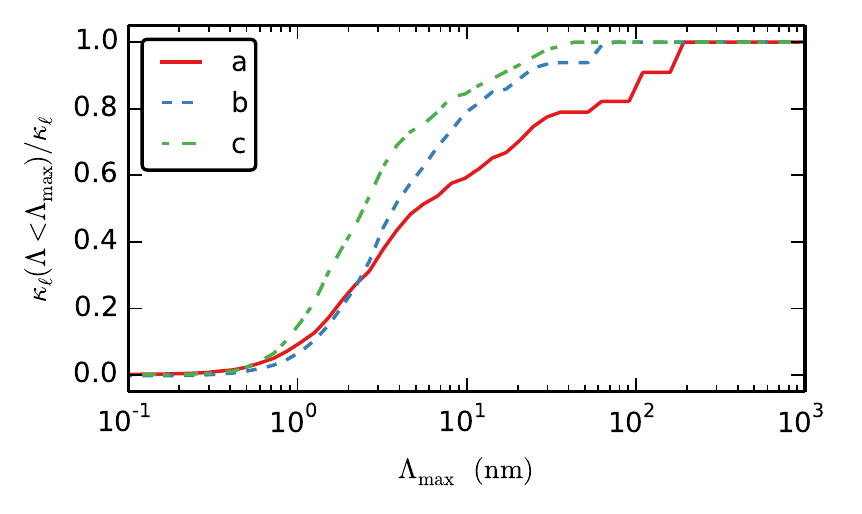}
  \end{center}
  \vspace{-5mm}
  \caption{Normalized cumulative lattice thermal conductivity of the Pnma phase of SnSe at $T=500\,\mathrm{K}$.}
  \label{fig:cumulative}
\end{figure}

The polycrystalline character of a sample tends to shorten the phonon mean free paths (MFPs) below their intrinsic bulk values.
To address the consistency of our monocrystalline results with experimental polycrystalline measurements, we must examine the MFPs of the phonons responsible for thermal transport. 
In Fig. \ref{fig:cumulative} we show the reduction in $\kappa_{\ell}$ obtained by removing the contribution from all phonons with MFPs above a certain threshold, and at an intermediate temperature $T=500\,\mathrm{K}$.
All relevant MFPs lie below $100\,\mathrm{nm}$.
It is therefore relatively safe to compare the thermal conductivity of a single crystal to those of polycrystalline samples with grain sizes on the order of $10$ to $100\,\mathrm{\mu m}$, such as those recently reported.\cite{Snyder_JMCA_SnSe_2014}

However, this comparison cannot be fully quantitative.
One reason is due to our lack of a detailed description of grain boundaries in each particular experimental sample.
More importantly, the methodology used in Refs. \onlinecite{Lenoir_APL_SnSe_2014} and \onlinecite{Snyder_JMCA_SnSe_2014} cannot fully resolve the anisotropy of the material.
Measurements are provided only along two directions: parallel and perpendicular to the pressing direction.
Based on the experimental details and on x-ray powder diffraction patterns,\cite{Lenoir_APL_SnSe_2014} perpendicular measurements are expected to represent a linear combination of $\kappa_{\ell}^{b}$ and $\kappa_{\ell}^{c}$.
Meanwhile, parallel measurements can include contributions from all three axes. 

All things considered, the comparison between single-crystal theoretical calculations and the experimental results from polycrystals shown in Fig. \ref{fig:kappa} is satisfactory.
More specifically, these results demand that at least one of the three principal values of the bulk thermal conductivity be significantly higher than any of those provided in Ref. \onlinecite{Kanatzidis_Nature_SnSe_2014}.
 Recalculating the figure of merit reported in Ref. \onlinecite{Kanatzidis_Nature_SnSe_2014} using our $\kappa_{\ell}$ brings it more in line with the largely isotropic values measured in polycrystals.\cite{Lenoir_APL_SnSe_2014,Snyder_JMCA_SnSe_2014}

In conclusion, we have used a fully \textit{ab-initio} method to study the lattice thermal conductivity of the Pnma phase of SnSe over a broad temperature range.
  We were motivated by three different sets of experimental results.
  One was performed on monocrystalline samples,\cite{Kanatzidis_Nature_SnSe_2014} and the remaining two on polycrystals.\cite{Lenoir_APL_SnSe_2014,Snyder_JMCA_SnSe_2014}
  The latter two studies seem to contradict the former, which would place SnSe as the best thermoelectric ever measured.
  Our results for the long axis of the SnSe structure agree with Ref. \onlinecite{Kanatzidis_Nature_SnSe_2014}.
  Over the other two axes our calculations differ from those measurements in two crucial aspects.
  One of these is that the computed values of $\kappa_{\ell}$, while still remarkably low, are higher than those measured.
  The other is that the two short axes are completely non-equivalent from a thermal transport viewpoint.
  In contrast, the limited amount of information about monocrystalline SnSe that can be extracted from the experiments on polycrystals \cite{Lenoir_APL_SnSe_2014,Snyder_JMCA_SnSe_2014}
  seems to be fully compatible with our results, as are historical measurements on single crystals.\cite{Wasscher_maxZT_SnS_SnSe_SSE_1963}

Our calculations have allowed us to identify a previously unreported in-plane anisotropy between the $b$ and $c$ axes.
  Our detailed study of the phonon frequencies, group velocities, and mean free paths help illuminate the structural foundations for such behavior.
  This is consistent with a thermoelectric figure of merit with a lower degree of anisotropy, as observed in polycrystals.\cite{Lenoir_APL_SnSe_2014,Snyder_JMCA_SnSe_2014}
  The information provided by our study will hopefully be useful for designing improved nanoengineering approaches to further optimize this promising thermoelectric.
  
  \begin{acknowledgments}
  The authors thank Dr. E. Alleno, Dr. G. Bernard-Granger, Dr. J. Heremans, Dr. B. Lenoir and Dr. A. Stelling for insightful discussion.
  This work is partially supported by the French ``Carnot'' project SIEVE, by DOD-ONR (N00014-13-1-0635, N00014-11-1-0136, N00014-09-1-0921).
  S. C. acknowledges partial support by DOE (DE-AC02-05CH11231), specifically the BES program under Grant \#EDCBEE.
  The consortium {\small AFLOWLIB}.org acknowledges Duke University --- Center for Materials Genomics --- and the CRAY corporation for computational assistance.
\end{acknowledgments}

\end{document}